\documentclass{llncs} 
 \usepackage{epsfig}  

\usepackage{amssymb} 
\usepackage{latexsym} 
 
\begin{document} 
\mainmatter 
 
\pagestyle{plain} 
\bibliographystyle{plain} 
 
\title{CLP versus LS on \\
 Log-based Reconciliation Problems}
 
\author{Fran\c{c}ois Fages} 
 
\authorrunning{Fran\c{c}ois Fages (INRIA)} 
 
\institute{ 
Projet Contraintes, INRIA-Rocquencourt,\\
BP105, 78153 Le Chesnay Cedex, France, \\ 
\email{francois.fages@inria.fr}\\
This article was presented at the 6th workshop of the \\
ERCIM WG on Constraints (Prague, June 2001).
Revised August 2001.} 
 
\def\implies{\Rightarrow}
 
\maketitle 
\begin{abstract} 
Nomadic applications create replicas of shared objects that evolve
independently while they are disconnected. When reconnecting, the
system has to reconcile the divergent replicas.  In the log-based
approach to reconciliation, such as in the IceCube system, 
the input is a common initial state and
logs of actions that were performed on each replica.  The output is a
consistent global schedule that maximises the number of accepted
actions. The reconciler merges the logs according to the schedule, and
replays the operations in the merged log against the initial state,
yielding to a reconciled common final state.

In this paper, we show the NP-completeness of the log-based reconciliation problem
and present two programs for solving it. Firstly, a constraint logic program (CLP) that uses
integer constraints for expressing precedence constraints, boolean constraints
for expressing dependencies between actions, and some heuristics for guiding the search.
Secondly, a stochastic local search method with Tabu heuristic (LS), 
that computes solutions in an incremental fashion
but does not prove optimality.
One difficulty in the LS modeling lies in the handling of both boolean variables
and integer variables, and in the handling of the objective function
which differs from a max-CSP problem.
Preliminary evaluation results indicate better performance for the CLP
program which, on somewhat realistic benchmarks, finds nearly optimal solutions
up to a thousands of actions and proves optimality up to a hundreds of
actions. 
\end{abstract}

\section{Introduction}

Data replication is a standard technique in distributed systems
to make data available in different sites. The different sites may be disconnected 
(mobile computing) or 
connected (groupware) in which case shared data are replicated for efficiency reasons
in order to avoid access through the network.
Obviously the replication of mutable shared data may cause conflicts,
the replicas may diverge into inconsistent states that have to be reconciled.
Nomadic applications create replicas of shared objects that evolve 
independently while they are disconnected, when reconnecting
the system has to reconcile the divergent replicas. 

What constitutes a conflict depends on the semantics of the application
and on the user's intent.
For example, in the version management system CVS \cite{CVS},
write actions are said to conflict if and only if they occur
in the same line of the same text file.
Accordingly, many existing reconcilers are restricted to specific
data types, such as source files, or file systems \cite{BP98}
or calendars.
In contrast, in the log-based approach to reconciliation \cite{PSTTS97,EMPSTT97,KRSD01}, 
best examplified in the IceCube system \cite{KRSD01},
the input is a common initial state and logs of actions
that were performed on each replica.
In this setting, an action is composed of a precondition, an operation and a postcondition
\cite{SRK00}.
The output is a consistent global schedule that maximises 
the number of accepted actions. The reconciler merges the logs according to the schedule,
and replays the operations in the merged log against the initial state,
yielding to a reconciled common final state \cite{KRSD01}.

In this paper, we show the NP-hardness of this reconciliation problem,
by encoding SAT as a reconciliation problem, and study two programs for solving it.
In section \ref{clp} we present a constraint logic program (CLP) that uses
boolean constraints
for expressing dependencies between actions, 
constraints over integers for expressing precedence constraints, 
and some heuristics for guiding the search during branch-and-bound optimization.
We evaluate the performance of this program
on a set of randomly generated benchmarks,
intended to modelize realistic log-based reconciliation problems,
with {\em densities} (defined
as the ratio between the number of constraints and the number of variables) 
1.5 for both precedence and dependency constraints between actions.
In these densities the CLP program finds quasi-optimal solutions up to a thousands of actions,
and proves optimality up to a hundreds of actions.

In section \ref{tabu} we present another program based on
a stochastic local search method with Tabu heuristics.
It computes solutions in an incremental fashion,
can use the initial logs of the application as starting solution,
but does not prove optimality.
One difficulty in the LS modeling lies in the handling of both boolean variables
and integer variables, and in the handling of the objective function
which differs from a max-CSP problem.
Preliminary evaluation results indicate better performance for the CLP program.

In the last section we present our conclusion and on-going work.

\subsection{Related work}

Log-based reconciliation is a new topic
for which few algorithms have been developed.
The only implementation we know of is the IceCube system reported in \cite{KRSD01}.
It is worth noting that the objective function of maximizing the number of accepted actions,
is different from maximizing the number of satisfied constraints.
For that reason, the modeling of log-based reconciliation as a max-CSP problem is inadequate.
This is also the main reason why in our second program based on local search,
the min-conflict heuristics \cite{MJPL92ai}
or the adaptive search method of \cite{Codognet00ercim}
do not perform well in our modeling, and we use instead a randomized Tabu heuristics.

In \cite{Fages01inria} we investigate the average-time complexity
of the CLP program as a function of both densities for
dependency and precedence constriants between actions.
We demonstrate the existence of a single computational
complexity peak on randomly generated problems, around densities 7 for
precedence constraints and 0 for dependency constraints between
actions. Around this peak we observe phase transitions
in the two dimensions of the density,
where the mean running time of the program shifts
from polynomial in the order to exponential.
These experimental investigation of the average-case complexity
of the CLP program are of quite general interest for 
the design of log-based reconciliation algorithms.
In particular they indicate where the hardest problems are,
and they clearly show that
it is crucial to use dependency constraints in an active way when computing
a schedule as these constraints greatly reduce the complexity of the problem.

\section{The log-based reconciliation problem}
\subsection{Statement of the optimization problem}
We have to reconcile a set of logs of actions
that have been realized independently, by trying to accept
the greatest number of actions as possible.
 
\paragraph{Input:}
A finite set of $L$ initial logs of actions $\{[T_i^1,...,T_i^{n_i}]\ |\ 1\le i\le L\}$,
some dependencies between actions $T_i^j \implies T_k^l$,
meaning that if $T_i^j$ is accepted then $T_k^l$ must be accepted,
and some precedence constraints $T_i^j<T_k^l$, meaning that if the actions are accepted 
they must be executed in that order. The precedence constraints
are supposed to be satisfied inside the initial logs.
 
\paragraph{Output:}
A subset of accepted actions, of maximal cardinality,
satisfying the dependency constraints,
given with a global schedule $T_i^j<...<T_k^l$ 
satisfying the precedence constraints.

Note that the output depends solely of the precedence constraints
between actions given in the input. In particular it is independent of the precise
structure of the initial logs.
The initial consistent logs can thus be used as starting solutions
in some algorithms but can be forgotten as well without affecting the output. 

\subsection{Complexity}

\begin{proposition}
The decision problem, i.e.~finding a schedule of a given length, is NP-complete,
even without dependency constraints.
\end{proposition}
\begin{proof}
The decision problem is obviously in NP. Indeed,
for any guessed schedule,
one can check in polynomial time whether the schedule is consistent.

NP-completeness is shown by encoding SAT into a reconciliation problem
with singleton initial logs and precedence constraints only.

Let us assume a SAT problem over $N$ boolean variables with $C$ clauses.
For each boolean variable $p$, we associate $2*C$ actions $p_0^1$, $p_1^1$, ..., $p_0^C$, $p_1^C$,
with precedence constraints $p_0^i < p_1^j$ and $p_1^j < p_0^i$ for all clause indices $i,j$ in $[1,C]$. 
The actions $p_0^i$ and $p_1^j$ are thus mutually exclusive for all clause indices $i,j$. 
We represent the valuation false for $p$ by accepting the actions $p_0^i$ for all $1\le i\le C$, and 
the valuation true by accepting $p_1^j$ for all $1\le i\le C$.
This defines a one-to-one mapping $\sigma$
between valuations over $N$ boolean variables,
and the accepted actions in schedules of length $N*C$ satisfying the mutual exclusion constraints.

For each clause, such as $p \vee q \vee\neg r$, we associate the precedence constraints
$p_0^i < q_0^i < r_1^i < p_0^i$ where $i$ is the index of the clause.
Being cyclic, these precedence constraints forbid to take simultaneously the actions
$p_0^i,\ q_0^i,\ r_1^i$ and $p_0^i$, that is, they encode the 
equivalent formula $\neg(\neg p \wedge \neg q \wedge r)$.
Hence a valuation $\eta$ satisfies a clause if and only if
the actions in $\sigma(\eta)$ satisfy all the precedence constraints associated
to the clause.
Note that, unlike the mutual exclusion constraints,
the precedence constraints are posted between action variables with the same clause index only.

Now we prove that a set of $C$ clauses over $N$ variables is satisfiable
if and only if there exists a schedule accepting $N*C$ actions and satisfying the
mutual exclusion constraints and the precedence constraints associated to the clauses.

The implication is clear: if $\eta$ is a valuation which satisfies all the clauses,
then $\sigma(\eta)$ is a set of $N*C$ actions which satisfies the mutual exclusion constraints,
and which can be ordered with increasing clause indices
and according to the precedence constraints for variables with the same clause index.

For the converse, let us suppose that we have a consistent schedule of $N*C$ actions.
Because of the mutual exclusion constraints, the schedule defines a valuation
of the SAT problem: indeed for each propositional variable $p$,
either $p_O^i$ is accepted for all $i$, and $p$ is false, either $p_1^i$ is accepted for all $i$,
and $p$ is true.
Furthermore the precedence constraints between actions of index $i$ establish
that that valuation satisfies the $i$th clause.
Therefore the valuation associated to the schedule satisfies all the clauses.
\hfill{QED.}
\end{proof}

\section{A CLP(FD,B) approach}\label{clp}

\subsection{Modeling with mixed boolean and integer constraints}

In this modeling of the problem, we forget the initial (consistent) logs of actions
and consider that all actions are at the same level.
We have $n$ elementary actions to which we associate:
\begin{itemize}
\item
  $n$ boolean variables $a_1,...,a_n$ which say whether the action is accepted or not
\item
  $n$ integer variables $p_1,...,p_n$ which give the position of the accepted actions
             in the global schedule
\end{itemize}

We have some dependency constraints 
$$a_i \implies a_j$$
and some precedence constraints     
$$a_i \wedge a_j \implies (p_i < p_j)$$
                    or equivalently, assuming false is 0 and true is 1,
$$a_i*a_j*p_i<p_j$$
We want to maximize $a_1+...+a_n$.

The search for solutions goes through an enumeration of the boolean variables $a_i$'s,
with the heuristics of instanciating first the variable $a_i$ which has
the greatest number of constraints on it (i.e.~first-fail principle
w.r.t.~the number of posted constraints)
and trying first the value 1 (i.e.~best-first search for the maximization problem).

This leads to the following straightforward CLP(FD,B) program (given in GNU-Prolog syntax):
       
{\small       
\begin{verbatim}                        
solve(Transactions,Dependencies,Precedences,Schedule) :-
        length(Transactions,N),
        length(La,N), fd_domain_bool(La),
        length(Lp,N), fd_domain(Lp,1,N),
        dependencies(Dependencies, Transactions, La),
        precedences(Precedences, Transactions, La, Lp),
        sum(La,S),
        fd_maximize(
           fd_labeling(La,[variable_method(most_constrained),
                           reorder(true),
                           value_method(max)]),
           S),
        fd_labeling(Lp,[value_method(min)]),
        schedule(La, Lp, Transactions, Keysort),
        sort(Keysort,Schedule).
 
dependencies([],_,_).
dependencies([(X#==>Y)|L],T,La):-
        nth(I,T,X), nth(I,La,A),
        nth(J,T,Y), nth(J,La,B),
        A#==>B,
        dependencies(L,T,La).
 
precedences([],_,_,_).
precedences([(X#<Y)|L],T,La,Lp):-
        nth(I,T,X), nth(I,La,A), nth(I,Lp,P),
        nth(J,T,Y), nth(J,La,B), nth(J,Lp,Q),
        A*B*P#<Q,
        precedences(L,T,La,Lp).
 
sum([],0).
sum([B|L],S):- S#=B+R, sum(L,R).
 
schedule([],[],[],[]).
schedule([B|La],[P|Lp],[T|Tr],S):-
        ((B=0)
        -> schedule(La,Lp,Tr,S)
        ;  S=[(P-T)|R], schedule(La,Lp,Tr,R)).                     
\end{verbatim}
}

Note that the labeling done in the optimization predicate
proceeds through the boolean variables only.
It is well known indeed that 
interval propagation algorithms provide indeed a complete procedure
for checking the satisfiability of precedence constraints.
It is thus not necessary to enumerate the possible values
of the position variables in the schedule,
as we know that the earliest dates are consistent.
The labeling of the positions
is done outside the optimization predicate,
just to compute a ground schedule
by taking the earliest dates for each action, without backtracking.

\subsection{Benchmarks}

At the present time, we do not have benchmarks of real-life log-based
reconciliation problems. Nevertheless we expect that in real-life reconciliation problems,
the optimal solutions accept more than 80\% of the actions typically.
These considerations, plus some preliminary inspections at some calendar applications
or the jigsaw problem presented in \cite{KRSD01},
lead us to create a benchmark of randomly generated problems
with density 1.5 for both precedence constraints
and dependency constraints.
We added a second series of more difficult benchmarks generated with
the same density 1.5 for precedence constraints
but without dependency constraints.

\begin{table}
\begin{center}
\small
\begin{tabular}{|c|c|c|c|c|c|c|c|c|c|}
\hline
\multicolumn{4}{|c|}{Benchmark} & \multicolumn{5}{|c|}{CLP} \\
\hline
Bench & Size & \#dep & \#prec & First sol. & Time & Opt. sol & Time & Proof \\
&&&&& (ms) && (ms) & (ms)\\
\hline
r100v1 & 100 & 120 & 151 & 98 & 60 & 98 & 110 & 110 \\
r100v2 & 100 &163 & 141 &75 & 80 & 77 & 90 & 190\\
r100v3 & 100 &139 & 145 &94 & 70 & 95 & 130 & 240\\
r100v4 & 100 &138 & 152 & 100 & 60 & 100 & 60 & 70\\
r100v5 & 100 &134 & 162 & 52 & 70 & 52 & 70 & 80\\
r200v1 & 200 & 315 & 284 &64 & 240 & 65 & 260 & 300 \\
r200v2 & 200 & 260 & 301 &189  & 270 & 191 & 28330 & 110600\\
r500v1 & 500 &740 & 784 & 198 & 1400 & 198 & 1400 & 1540\\
r500v2 & 500 & 761 & 736 & 490 & 1650 & $\ge$491& 65090 & ??\\
r800v1 & 800 & 1181 & 1165 & 770 & 4990 & ?? & ?? & ??\\
r800v2 & 800 & 1237 & 1134 & 318 & 3410 & 318 & 3410 & 5580\\
r1000v1 & 1000 & 1493 & 1562 & 389 & 5500 & 389 & 5500 & 6080\\
r1000v2 & 1000 & 1462 & 1586 & 938 & 42880 & ?? & ?? & ??\\
\hline
t40v1 & 40 & 0 & 63 & 36 & 10 & 36 & 10 & 340 \\
t40v2 & 40 & 0 & 64 & 36 & 10 & 37 & 180 & 440 \\
t40v3 & 40 & 0 & 49 & 38 & 10 & 38 & 10 & 20 \\
t40v4 & 40 & 0 & 64 & 35 & 10 & 37 & 110 & 240 \\
t50v1 & 50 & 0 & 82 & 44 & 20 & 45 & 1080 & 6220 \\
t50v2 & 50 & 0 & 78 & 46 & 10 & 47 & 240 & 630 \\
t50v3 & 50 & 0 & 84 & 43 & 10 & 44 & 970 & 8120 \\
t50v4 & 50 & 0 & 67 & 45 & 10 & 46 & 90 & 200 \\
t70v1 & 70 & 0 & 91 & 67 & 10 & 68 & 530 & 1120 \\
t70v2 & 70 & 0 & 104 & 67 & 20 & 67 & 20 & 190 \\
t80v1 & 80 & 0 & 120 & 75 & 30 & ?? & ?? & ??\\
t100v1 & 100 & 0 & 146 & 93 & 50 & ?? & ?? & ?? \\
t200v1 & 200 & 0 & 323 & 189 & 200 & ?? & ?? & ??\\
t500v1 & 500 & 0 & 740 & 482 & 1980 & ?? & ?? & ??\\
t800v1 & 800 & 0 & 1181 & 772 & 7230 & ?? & ?? & ??\\
t1000v1 & 1000 & 0 & 1459 & 954 & 29020 & ?? & ?? & ??\\
\hline
\end{tabular}
\end{center}
\caption{Experimental results.}\label{benchCLP}
\end{table}

Table \ref{benchCLP} depicts the experimental results on 
both series of benchmarks.
The size given in the second column is the total number of actions.
The numbers of dependency constraints and precedence constraints
are given in the following columns. These constraints are generated
for each pair of actions randomly, with probability $1.5/size$ 
which gives $1.5*size$ constraints of each type in average.
The second series of benchmarks contains no dependency constraints.

The running times of this section
have been measured in GNU-Prolog on a 866MHz Pentium III PC under Linux.
We indicate, in order, the number of accepted actions in the first 
schedule found, the running time for finding this solution,
the number of accepted actions in the optimal schedule,
the running time for finding the optimal schedule (from the start),
and finally the total running time including the proof of optimality.

The first solution found in these benchmarks 
is always very near the optimal solution.
This indicates that the most-constrained variable choice heuristics 
combined with the best-first search heuristics
perform very well in this modeling of the problem.
Optimal solutions with their optimality proof are always computed in less than a second for
problems with up to a hundreds of actions.
On larger problems optimality proofs become difficult to obtain
but the first solution found is always satisfying and fast to compute.
The problems without dependency constraints are harder to solve.
Optimality proofs are nevertheless always obtained on problems of size up to 50.

\section{A stochastic local search approach}\label{tabu}

Approximated solutions to hard optimization problems can be found 
with heuristics.
Local search is one of the fundamental heuristics that has been shown particularly
effective for many classes of applications,
including some classical benchmarks of constraint programming \cite{MVH99informs,Codognet00ercim,GH01aor}.
Local search methods iterate the transformation of some
initial solution $s_0$, by choosing at each step the next solution, $s_{i+1}$,
in some neighborhood $N(s_i)$ of $s_i$.
Descent methods choose at each iteration the neighbour
which minimizes the objective function $f$: 
$s_{i+1}=argmin_{s\in N(s_i)} f(s)$
and stops whenever $f(s_{i+1})>f(s_i)$.
Descent methods thus stop in the first encountered local optimum.
Multi-start methods iterate the application of the descent method
from different initial solutions, and stop with the best encountered
local optimum.
Simulated annealing, variable neighborhood methods and Tabu search are
local search methods that escape from local minima 
in an iterative fashion, without restarting descents.

\begin{figure}[t]
\begin{center}
\epsfig{file=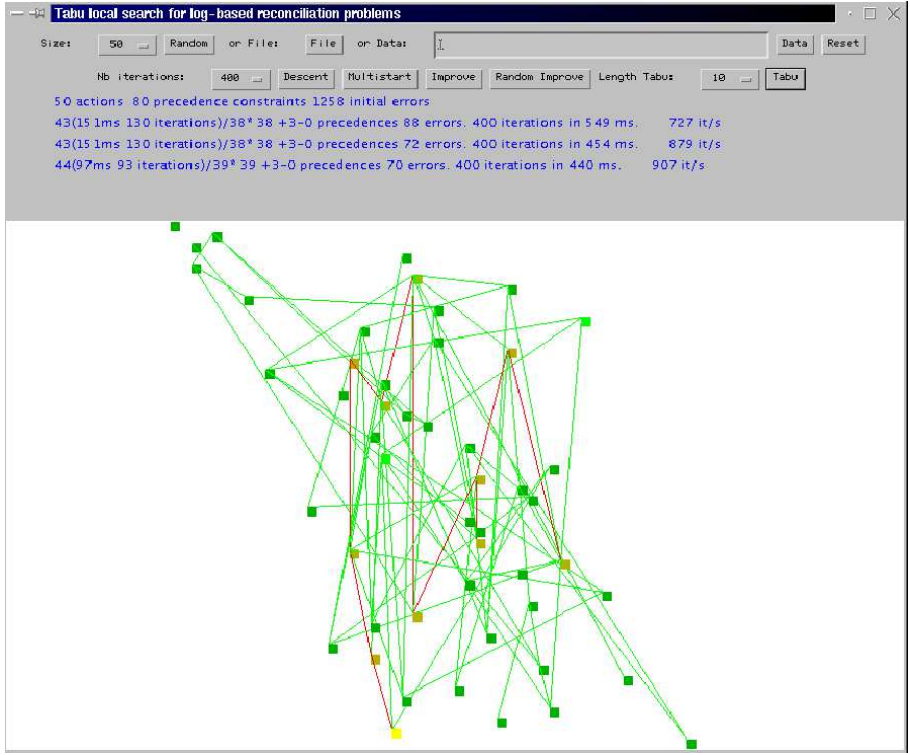, width=12cm, height=10cm} 
\end{center}
\caption{Screen dump of the local search program.}\label{applet}
\end{figure}

Tabu search \cite{GL98} consists in choosing at each step $i$
the state 
$$s_{i+1}=argmin_{s\in N(s_i)} f(s)$$
 even if $f(s_{i+1})$ is greater
than the best solution found. A Tabu list $L$ records the already visited
states. The Tabu states cannot be revisited, except if they improve
the objective function. Furthermore the Tabu list
is a short term memory, after some number of iterations, 
already visited states are deleted from the Tabu list and can be freely reconsidered.
In order to prevent cycles in the same set of solutions,
the size of the Tabu list, as well as the neighborhoods, can be changed dynamically
during the iterations.

The log-based reconciliation problem can be modeled quite naturally as a local search problem
for finding schedules that maximize the number of accepted actions.
But there is one difficulty to mix boolean variables and position variable,
and to define an appropriate evaluation function for guiding the search.

In this section we shall treat reconciliation problems with precedence constraints only.
One solution is to represent in the states the positions of the actions in the 
current schedule, and to count in the objective function the number of actions 
which have all their precedence constraints satisfied.
For guiding the search however, a more refined evaluation function is needed
in order to have a measure of progress towards better solutions
that do not yet improve the objective function.
We thus use an evaluation function that counts for each violated
precedence constraints 
$$p_i\not<p_j$$
the error
$$1+(p_i-p_j)$$
The local moves are simply the incrementation or the decrementation
by 1 of the position of an action in the schedule.

The min-conflict heuristic \cite{MJPL92ai} consists in choosing
for a move the variable with a highest error
and the move which minimizes that error.
Nevertheless in the reconciliation problem, improving the evaluation function
on the variables with the highest error does not necessarily leads
to a good solution w.r.t.~the objective function, as actions with high errors
can be simply not accepted.
Therefore we don't use the min-conflict heuristic
and perform local moves on all actions as long as they improve the evaluation function.
The {\em value of a configuration} w.r.t. the objective function
is the number of actions with 0 errors.
Note that this value is in fact a lower bound of the cost of the configuration
w.r.t. the objective function, as constraints with unaccepted actions should be ignored.
The {\em cost of a configuration} w.r.t. the objective function is thus computed 
as the number of actions remaining after successively removing
the actions with the greatest number of violated constraints with non removed actions.

This defines the descent method evaluated in the next section.
The Tabu search method adds an adaptive memory to
escape from local minima. When a local minimum is reached on a variable
it is marked in the Tabu list for a while, and is not reconsidered
except if it is the only way to improve the solution.
In order to increase the diversification,
the number of iterations for which the variables is marked Tabu
is randomly generated between 1 and some maximum Tabu length value (10 in the experiments).
Furthermore if a local minimum is reached on all variables,
one variable is chosen randomly to make a move that degrades the evaluation function.

The user interface of the program visualizes
the movements of the actions in the schedule during the search, see figure \ref{applet}.
The graphical interface represents each action on a different line.
The position of the action in the line indicates 
the scheduling of the action.
Precedence constraints are materialized by lines
between actions, those lines in green represent satisfied
precedence constraints, those in yellow are violated.

\subsection{Evaluation}

\begin{table}
\begin{center}
\begin{tabular}{|c|c|c|c||c|c|c||c|c|c|}
\hline
\multicolumn{4}{|c|}{Benchmark} & \multicolumn{3}{|c|}{Descent method} & \multicolumn{3}{|c|}{Tabu search} \\
\hline
Bench & Size & \#prec. & Opt. & Best & \#iter. & Time & Best & \#iter. & Time  \\
&&&& & & (ms) & &  & (ms)\\
\hline
t40v1 & 40 & 63 & 36 & 32 & 21 & 29 & 36 & 811 & 935\\
t40v2 & 40 & 64 & 37 & 31 & 23 & 39 & 36 & 1126 & 1469\\
t40v3 & 40 & 49 & 38 & 34 & 20 & 28 & 38 & 1133 & 1307\\
t40v4 & 40 & 64 & 37 & 35 & 23 & 28 & 36 & 259 & 176 \\
t50v1 & 50 & 82 & 45 & 40 & 25 & 60 & 44 & 635 & 1027\\
t50v2 & 50 & 78 & 47 & 42 & 26 & 68 & 46 & 2028 & 3908\\
t50v3 & 50 & 84 & 44 & 32 & 25 & 61 & 40 & 1303 & 2351\\
t50v4 & 50 & 67 & 46 & 42 & 26 & 55 & 46 & 727 & 1253\\
t70v1 & 70 & 91 & 68 & 59 & 36 & 153 & 65 & 2375 & 8980\\
t70v2 & 70 & 104 & 67 & 59 & 46 & 212 & 67 & 1325 & 5026\\
t80v1 & 80 & 116 & 77 & 61 & 43 & 299 & 70 & 3079 & 15523\\
%t80v2 & 80 & 120 & $\ge$75 & & & & & & \\
t100v1 & 100 & 155 & $\ge$91 & 79 & 66 & 1265 & 90 & 5426 & 41524\\
t200v1 & 200 & 323 & $\ge$189 & 164 & 125 & 17977 & 181 & 10461 & 357712\\
t500v1 & 500 & 740 & $\ge$482 & 400 & 198 & 890990 & ?? & ?? & ?? \\
t800v1 & 800 & 1181 & $\ge$772 & 680 & 595 & 7450661 & ?? & ?? & ?? \\
t1000v1 & 1000 & 1459 & $\ge$954 & 840 & 523 &  14543805& ?? & ?? & ?? \\
\hline
\end{tabular}
\end{center}
\caption{Experimental results (benchmarks without dependency constraints).}\label{benchTabu}
\end{table}

The local search algorithm described in the previous section
has been implemented in Java.
Table \ref{benchTabu} depicts the performance results 
on the previous series of benchmarks without dependency constraints.
The running times have been measured with the Java 2 v1.3 JDK compiler.
For each bench we recall the number of accepted actions
in the optimal solution found by the CLP program,
and present the best solution found by the (deterministic) descent method,
and the best solution found by a run of the (randomized) Tabu search method.

These results indicate that the descent method leads to local minima 
of poor quality, and that the Tabu search program succeeds
in escaping from these local minima.
However on large instances, the convergence to better solutions is very long.

Better tuning of the Tabu search method might still improve the 
performance of the method in these benchmarks, in particular by improving
the diversification strategies.
Experimental results obtained with the min-conflict heuristic
gave worse results for the reasons explained in the previous section.
A different modeling seems necessary to use these heuristics,
improve the handling of the objective function
and treat dependency constraints.

\section{Conclusion}

Reconciliation problems in nomadic applications present interesting combinatorial
optimization problems, with both static and on-line versions.
We have studied an NP-hard log-based reconciliation problem.
Its modeling with boolean and precedence constraints
lead to a straightforward CLP(FD,B) program that finds
nearly optimal solutions up to a thousands of actions,
and proves optimality up to a hundreds of actions on realistic benchmarks.
Enumeration proceeds through the boolean variables only.
The precedence constraints are propagated in a complete way
which makes enumeration superfluous.
This program could still benefit however from more efficient algorithms
for detecting cycles in precedence constraints,
with a complexity independent of the size of the variables' domain \cite{DMP91}.

We have developed a second program based on local search with Tabu heuristic.
One potential advantage for the local search approach is that it can
benefit from the initial logs as starting solution,
and can provide solutions incrementally for the on-line reconciliation problem.
Currently this program performs however poorly in comparison to the CLP program.
One difficulty lies in the handling of both boolean variables
and integer variables which represent the positions of the
actions in the schedule. 
In our modeling of the problem, the adaptive search method of \cite{Codognet00ercim}
did not perform well because of the difficulty to handle the objective function
which differs from a max-CSP problem.
Other modelings are thus currently under investigation.

\vspace{5mm}
\noindent
{\bf Acknowledgment.}
I would like to thank Silvano Dal Zilio,  Peter Dreuschel, C\'edric Fournet,
Anne-Marie Kermarrec, Marc Shapiro and Antony Rowstron
for fruitful discussions on this application.
I am grateful also to Philippe Codognet, Daniel Diaz and Charlotte Truchet
for tabu-free discussions about search methods.

%\nocite*
\bibliography{biblio}

\begin{thebibliography}{10}

\bibitem{BP98}
S.~Balasubramanian and B.~Pierce.
\newblock What is a file synchronizer?
\newblock In {\em Proc ACM/IEEE Int. Conf on Mobile Computing and Networking},
  1998.

\bibitem{Codognet00ercim}
P.~Codognet.
\newblock Adaptive search: preliminary results.
\newblock In {\em Proc. of fourth ERCIM/CompulogNet Workshop on Constraints},
  Venice, Italy, june 2000.

\bibitem{DMP91}
R.~Dechter, I.~Mieri, and J.~Pearl.
\newblock Temporal constraint networks.
\newblock {\em Artificial Intelligence}, 49(1-3):61--95, january 1991.

\bibitem{EMPSTT97}
W.K. Edwards, E.D. Mynatt, K.~Petersen, M.J. Spreitzer, D.B. Terry, and M.M.
  Theimer.
\newblock Designing and implementing asynchronous applications with bayou.
\newblock In {\em Proc. Int. Symp. on User Int. Software and Tech.}, Alberta,
  Canada, october 1997.

\bibitem{CVS}
P.~Cederqvist et~al.
\newblock Version management with cvs, 1992.

\bibitem{Fages01inria}
F.~Fages.
\newblock A constraint programming approach to log-based reconciliation
  problems for nomadic applications.
\newblock Technical report, INRIA Rocquencourt, France, May 2001.

\bibitem{GH01aor}
P.~Galinier and J.K. Hao.
\newblock A general approach for constraint solving by local search.
\newblock {\em Submitted to Annals of Operations Research}, january 2001.

\bibitem{GL98}
F.~Glover and M.~Laguna.
\newblock {\em Tabu search}.
\newblock Kluwer Academic Publishers, 1998.

\bibitem{KRSD01}
A.M. Kermarrec, A.~Rowstron, M.~Shapiro, and P.~Druschel.
\newblock The {I}ce{C}ube approach to the reconciliation of divergent replicas.
\newblock In {\em Proc. of Twentieth ACM Symposium on Principles of Distributed
  Computing PODC}, Newport, RI USA, August 2001.

\bibitem{MVH99informs}
L.~Michel and P.~Van Hentenryck.
\newblock Localizer: A modeling language for local search.
\newblock {\em INFORMS Journal of Computing}, 11(1):1--14, 1999.

\bibitem{MJPL92ai}
S.~Minton, M.~Johnson, A.~Philips, and P.~Laird.
\newblock Minimizing conflicts: a heuristic repair method for constraint
  satisfaction and scheduling problems.
\newblock {\em Artificial Intelligence}, 58:161--205, 1992.

\bibitem{PSTTS97}
K.~Petersen, M.J. Spreitzer, D.B. Terry, M.M. Theimer, and D.J. Demers.
\newblock Flexible update propagation for weakly consistent replication.
\newblock In {\em Proc. ACM SIGOPS Symp. on Operating Systems Principles
  SOSP-16}, pages 288--301, Saint-Malo, France, 1997.

\bibitem{SRK00}
M.~Shapiro, A.~Rowstron, and A.M. Kermarrec.
\newblock Application-independent reconciliation for nomadic application.
\newblock In {\em Proc. of SIGOPS European workshop Beyond the PC: new
  challenges for the operating system}, Kolding, Danemark, september 2000.

\end{thebibliography}

\end{document}